\documentclass[pre,amsmath,amssymb,twocolumn,superscriptaddress]{revtex4-2}

\usepackage{amsmath,amssymb}
\usepackage[usenames]{color}
\usepackage{amssymb}
\usepackage{dsfont}
\usepackage{grffile}
\usepackage{textcomp}
\usepackage{xspace}
\usepackage{hyperref}
\usepackage{graphicx} 
\usepackage{outlines}

\newcommand{\tr}{{\rm tr}}

\newcommand{\im}{{\rm i}}

\newcommand{\sx}{\sigma^x}
\newcommand{\sy}{\sigma^y}
\newcommand{\sz}{\sigma^z}

\newcommand{\He}{\bar{H}}

\newcommand{\V}{\mathcal{V}}
\newcommand{\Ss}{\mathcal{S}}
\newcommand{\vlr}{v_{\rm LR}}

\newcommand{\sutd}{Science, Mathematics and Technology Cluster, Singapore
University of Technology and Design, 8 Somapah Road, 487372 Singapore}
\newcommand{\sutdepd}{EPD Pillar, Singapore University of Technology and Design, 8 Somapah Road, 487372 Singapore}
\newcommand{\como}{Center for Nonlinear and Complex Systems, Dipartimento
di Scienza e Alta Tecnologia, Universit\`a degli Studi dell'Insubria, via
Valleggio 11, 22100 Como, Italy}
\newcommand{\infn}{Istituto Nazionale di Fisica Nucleare, Sezione di Milano, via Celoria 16, 20133 Milano, Italy}
\newcommand{\nest}{NEST, Istituto Nanoscienze-CNR, I-56126 Pisa, Italy}
\newcommand{\riogrande}{International Institute of Physics, Federal University of Rio Grande do Norte, Natal, Brazil}

\begin{document}

\title{From ETH to algebraic relaxation of OTOCs in systems with conserved quantities}

\author{Vinitha Balachandran}
\affiliation{\sutd}
\author{Giuliano Benenti}
\affiliation{\como}
\affiliation{\infn}
\affiliation{\nest}
\author{Giulio Casati}
\affiliation{\como}
\affiliation{\riogrande}
\author{Dario Poletti}
\affiliation{\sutd}
\affiliation{\sutdepd}

\begin{abstract}
The relaxation of out-of-time-ordered correlators (OTOCs) has been studied as a mean to characterize the scrambling properties of a quantum system.
We show that the presence of local conserved quantities typically results in, at the fastest, an algebraic relaxation of the OTOC provided (i) the dynamics is local and (ii) the system follows the eigenstate thermalization hypothesis.
Our result relies on the algebraic scaling of the infinite-time value of OTOCs with system size, which is typical in thermalizing systems with local conserved quantities, and on the existence of finite speed of propagation of correlations for finite-range-interaction systems.
We show that time-independence of the Hamiltonian is not necessary as the above conditions (i) and (ii) can occur in time-dependent systems, both periodic or aperiodic.
We also remark that our result can be extended to systems with power-law interactions.
\end{abstract}

\maketitle

\setcounter{secnumdepth}{2}

The characterization of the long-time dynamics of quantum many-body isolated systems has been a focus of interest for many years \cite{Borgonovi_2016,D_Alessio_2016,Polkovnikov2011,Nandkishore_2015}. Important insights are provided by the eigenstate thermalization hypothesis (ETH) which states that eigenstates of a Hamiltonian which are close-by in energy have very similar local properties \cite{Deutsch,Srednicki,Rigol2008,D_Alessio_2016}.
A quantity that has shown to give further insights into the dynamics of quantum systems is the out-of-time-order correlator (OTOC), first introduced in \cite{Larkin1969}.
Given two operators $A$ and $B$, the OTOC is given by $\langle [A(t),B][A(t),B]^\dagger\rangle$, where $A(t)$ is the time-evolved $A$, $[\cdot,\cdot]$ is the commutator and the expectation value $\langle\cdot\rangle$ is taken over a suitably chosen state.

For quantum systems with a classical correspondence, one can choose $A$ and $B$ to be respectively the position $x$ and momentum $p$ operators, thus showing that the OTOC would grow exponentially in time at a rate given by the Lyapunov exponent \cite{Galitski2017,Hashimoto_2017,Cotler_2018,Ignacio2018,Ch_vez_Carlos_2019,Fortes2019, Rammensee2018,Prakash2020,Bergamasco2019,Pilatowsky2020,Xu2020,Rozenbaum2020, Wang2020,wang2020quantum}.
The fast scrambling of information in black holes has also been related, thanks to holography \cite{Witten1998, Maldacena1999}, to the process of thermalization in strongly interacting systems \cite{HaydenPreskill2007, SekinoSussking2008, Shenker_2014, SachdevYe1993, Kitaev, LashkariHayden2013,  RobertsStanford2015, CotlerTezuka2017, RobertsSussking2015, HosurYoshida2016, Borgonovi2019}.
In Ref.~\cite{Maldacena_2016} it has been conjectured that the growth rate of the OTOC can be bounded by the temperature. OTOCs have been evaluated in different experimental setups \cite{LiDu2017, GarttnerRey2017, LandsmanMonroe2019, JoshiRoos2020, BlokSiddiqi2021, MiYu2021, Jochen2021}.
A remarkable insight into the time evolution of OTOCs was achieved in a series of works in which it was found that OTOCs follow a ``hydrodynamic'' equation of motion in local random unitary circuit models \cite{NahumHaah2017, NahumHaah2018, KeyserlingkSondhi2018, RakovszkyKeyserlingk2018, KhemaniHuse2018}.

\begin{figure}[h]
\includegraphics[width=\columnwidth]{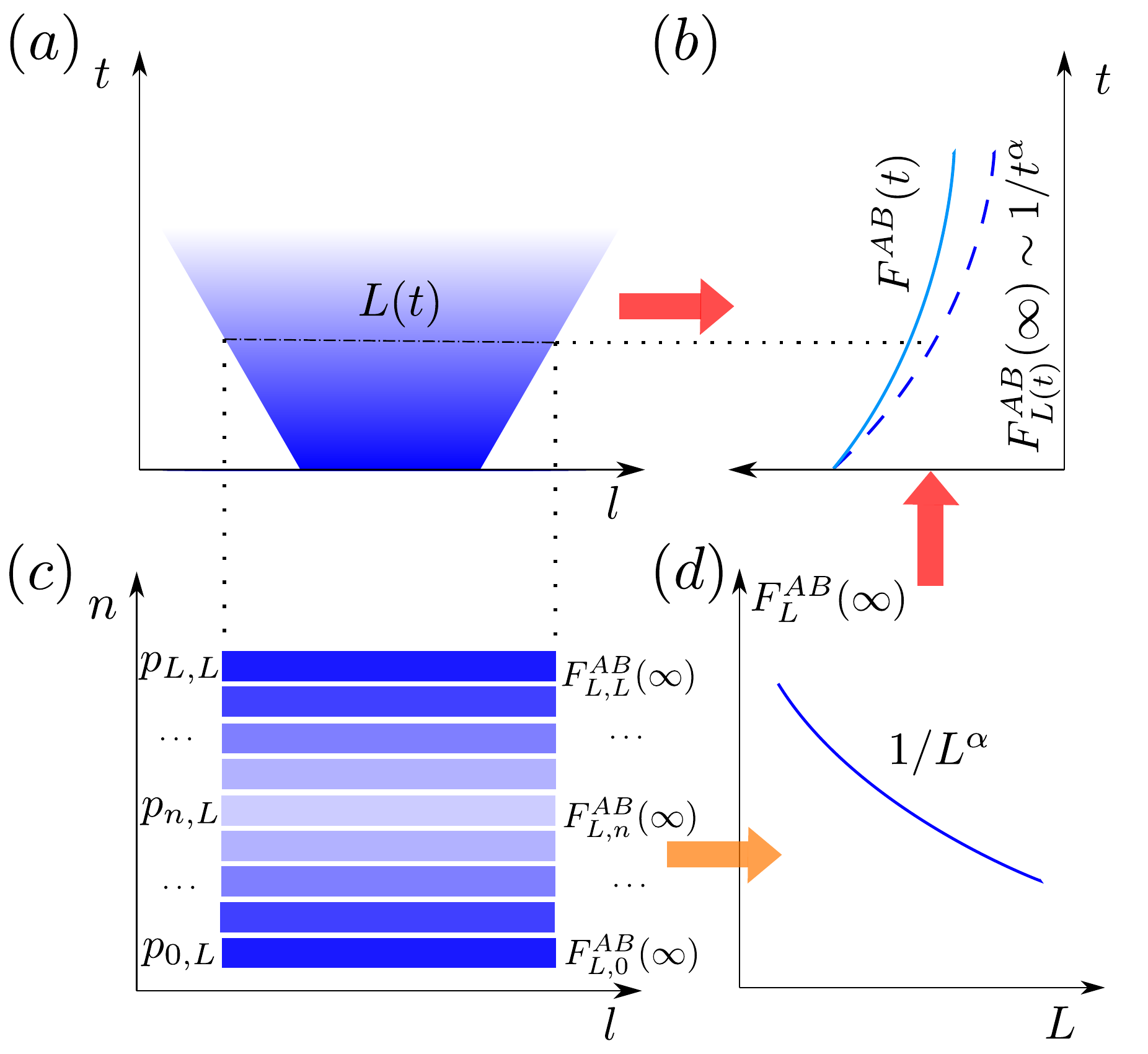}
\caption{(a) Depiction of spreading of correlation over a region $L(t)$ expanding as a light-cone. (b) Algebraic relaxation of an OTOC of operators $A$ and $B$, $F^{AB}(t)$, lower-bounded by the infinite-time value of the OTOC for a system of size $L(t)$. (c) Depiction of the different infinite-time values of the OTOC in each number sector $n$, $F^{AB}_{L,n}(\infty)$, each weighted with a different probability $p_{n,L}$. (d) Algebraic scaling of the infinite-time OTOC $F^{AB}(\infty)$ with the system size $L$.}
\label{fig:FigSummary}
\end{figure}

In this work, we investigate the emergence of slow relaxation of OTOCs in the presence of local conserved quantities.
We first show that if the infinite-time value of an OTOC scales algebraically with the system size, and the Hamiltonian is local, then it results that the OTOC relaxes (at the fastest) algebraically in time. This is pictorially depicted in Fig.\ref{fig:FigSummary} where the spreading of correlations in a local Hamiltonian, Fig.\ref{fig:FigSummary}(a), and algebraic scaling with system size of the infinite-time OTOC, Fig.\ref{fig:FigSummary}(d), imply a lower bound in the relaxation dynamics of the OTOC versus time which is algebraic, Fig.\ref{fig:FigSummary}(b). A more detailed description of Fig.\ref{fig:FigSummary} will be presented in Sec.\ref{sec:mechanism} and \ref{sec:time_ind_num_cons}.

We apply this result to both time-independent and time-dependent systems. We first consider number-conserving systems (where for us the quantum number is the total magnetization) and  provide estimates based on ETH for infinite-time value of the finite-size OTOC. Later we show that
one can expect a slow relaxation of the OTOC from the fact that the evolution up to a certain time of a thermodynamically large system is accurately emulated by that of an ensemble of initial conditions in different number sectors, with a weight given to each sector that depends on the system size. This, hence, results in a finite value of an OTOC which scales algebraically with the system size and, given the finite speed of propagation of correlations, implies a slow relaxation of the OTOC in time.
For local time-dependent Hamiltonians, we can also demonstrate that number conservation results in slow relaxation.
We will in fact show that this can be inferred by the fact that each number sector relaxes to a different infinite-time value, thus resulting in an finite-size infinite-time OTOC that scales algebraically with the system size, and by the fact that the propagation of correlation is still limited by the locality of the Hamiltonian.

For local time-independent systems with no local conserved quantities other than energy, it is also possible to show that OTOC relaxes algebraically because the infinite-time OTOC typically scales algebraically with the system size \cite{HuangZhang2019}.
Furthermore, the requirement of locality of the Hamiltonian can be relaxed considering systems with power-law couplings. This is possible because, in these systems, correlations propagate at an average speed that increases algebraically with time \cite{Foss-Feig2015, LuitzBarLev2019, ColmenarezLuitz2020, Kuwahara2020, Else2020}.

Our results are in agreement with those obtained in \cite{RakovszkyKeyserlingk2018, KhemaniHuse2018}, where the algebraic relaxation of the OTOC was derived for local random unitary circuit models, assuming that the OTOC dynamics is described by hydrodynamic equations. Here we show that the slow relaxation, obtained for random circuits, emerges in local systems like non-integrable spin chains without any randomness, as long as they follow ETH.

We corroborate our results with simulations of large non-integrable spin systems (with sizes up to $64$ spins).
To achieve such large system sizes we use a matrix product states algorithm (MPS) \cite{White1992, WhiteFeiguin2004, Daley_2004, Vidal2004, Schollwock_2011} and we sample over a large number of different initial conditions. For a detailed study on the effectiveness of MPS-related methods for the study of OTOC see \cite{HemeryLuitz2019}. 

This paper is organized in the following manner: In Sec.\ref{sec:infinite_time} we give a preliminary introduction to OTOCs and provide the reader with useful definitions. The infinite-time value of an OTOC is discussed in Sec.\ref{sec:estimate_con}. In Sec.\ref{sec:mechanism} we show that if the finite-size infinite-time OTOC of a local Hamiltonian scales algebraically with the system size, then the OTOC also relaxes, at the fastest, algebraically. In Sec.\ref{sec:time_ind_num_cons} we consider number-conserving  systems, both time-independent and time-dependent ones, and show that number conservation allows to predict an algebraic scaling of the finite-size infinite-time OTOC with the system size, from which we can deduce a slow relaxation of the OTOC in time. In Sec.\ref{sec:time_ind_non_num_cons} we show how, for local time-independent systems, one can infer that the long-time, finite-size, OTOC scales algebraically, even without local conserved quantities other than energy. This would thus result in a slow relaxation of the OTOC.
In Sec.\ref{sec:conclusions} we draw our conclusions, while in the Appendix \ref{app:numerical_details} we present some technical details for the evaluation of the OTOC on large systems, and discuss the evolution of the variance of the OTOC depending upon whether the decay is fast or slow.

\section{Definitions and method} \label{sec:infinite_time}
Considering two operators $A$ and $B$ the infinite-temperature out-of-time-ordered correlator (OTOC) is defined as
\begin{align}\label{otoc}
 O^{AB}(t) &= \langle [A(t), B][A(t), B]^\dagger\rangle
\end{align}
where $A(t) = U^{\dagger}A U$ is the time evolved operator $A$ due to the unitary evolution $U=\mathcal{T}e^{-\im \int_0^t H(\tau) d\tau}$ from the time-ordered integration of the (generically) time-dependent Hamiltonian $H(t)$. With the symbol $\langle\dots\rangle = \tr(\dots)/\mathcal{V}$ we indicate the average over the infinite-temperature state, where $\mathcal{V}$ is the size of the relevant Hilbert space. In the following, we will only consider operators $A$ and $B$ which are local, unitary and traceless. We note here that we imply traceless within the full Hilbert space, although the operators may have a non-zero trace within a number sector.

The OTOC can then be written as the sum of a time-ordered and a non-time-ordered part $O^{AB}(t) = 2G^{AB}(t)  - 2F^{AB}(t)$ where
\begin{align}\label{otoc_F}
 F^{AB}(t) &= \langle A(t)B A(t) B\rangle
\end{align}
while $G^{AB}=1$ because $A$ and $B$ are unitary.
Thus, in this case, interesting dynamics can only stem from the non-time-ordered component $F^{AB}(t)$.
For this reason, when we refer to slow decay of the OTOC
we mean the slow decay of the non-time-ordered part $F^{AB}(t)$.

In the following, calculations of the OTOCs are done with exact diagonalization methods for system with Hilbert space size of at most $100000$, while for large systems we use our matrix product states (MPS) algorithm \cite{White1992, WhiteFeiguin2004, Daley_2004, Vidal2004, Schollwock_2011, HemeryLuitz2019}.
For a given initial state $|\psi\rangle$, we compute the OTOC by calculating the overlap between the two evolved functions $F^{AB}(t)=\langle \psi_2|\psi_1\rangle$, where $|\psi_1\rangle= U^\dagger A U B |\psi\rangle$ , $|\psi_2\rangle= B U^\dagger A U  |\psi \rangle $.
One important aspect needs to be pointed out here though. To evaluate an infinite temperature OTOC, one needs to take an average over all basis states with equal probability. While this is readily possible for spin systems up to $L=20$, it starts to become more demanding for larger system sizes. In particular, for systems which we can only simulate using MPS, e.g. $L=64$, we cannot simulate all the possible states. Instead, we sample over a number of different initial conditions and average over them.
For more details see Appendix~\ref{app:numerical_details}.

For periodically driven systems with Hamiltonian $H(t)=H(t+T)$ with period $T$, can be stroboscopically emulated by that of an effective Floquet Hamiltonian $\He$ defined as
\begin{align}
U_T=\mathcal{T}e^{-\im \int_0^T H(\tau) d\tau} = e^{-\im \He T}. \label{eq:He}
\end{align}
We stress here that $\He$ always exists \cite{Fernandez1990, Casas2007, BlanesRos2008}, and we are not considering whether it can be obtained from a Magnus expansion or not. We refer to the eigenenergies and eigenstates of the effective Hamiltonian $\He$ as, respectively, Floquet eigenenergies and Floquet eigenstates.

When studying systems with a time-dependent but not periodic Hamiltonian $H(t)$, we can consider that the evolution up to this time $t$ was due to a time-independent Hamiltonian $\He_t$
\begin{align}
U_{0}^{t} = e^{-\im\int_0^t H(\tau)d\tau} = e^{-\im\He_t t}. \label{eq:Het}
\end{align}
where we have used the sub-index $t$ to indicate that $\He_t$ is the time-independent Hamiltonian relevant to describe the evolution up to a time $t$.

\section{Infinite-time value of OTOC}\label{sec:estimate_con}
Considering a time-independent Hamiltonian $H$ with eigenvalues $E_p$, the time evolution of the OTOC $F^{AB}(t)$ can always be written in the basis of the energy eigenstates $|p\rangle$ as
\begin{align}\label{otoc_energybasis}
 F^{AB}(t)= \frac{1}{\mathcal{V}}\sum_{p,q,k,l}e^{-\im(E_p-E_q+E_k-E_l)t}A_{pq}B_{qk}A_{kl}B_{lp}
\end{align}
where $A_{pq}=\langle p|A|q \rangle$ and $B_{qk}=\langle q|B|k \rangle$. We work in units for which $\hbar=1$. For time-dependent systems, the eigenstates and eigenvalues $|p\rangle$ and $E_p$ can be evaluated from the effective Hamiltonians $\He$ and $\He_t$ from Eqs.(\ref{eq:He},\ref{eq:Het}) depending on whether the system is periodic or not.

The only relevant terms at $t\rightarrow\infty$ in the above expression are those for which $E_p-E_q+E_k-E_l=0$. Considering a generic spectrum \cite{Srednicki_1998, HuangZhang2019}, the contributions to the infinite-time value of $F^{AB}$ are given by
\begin{align}\label{eq:otocdiag}
 F^{AB}(\infty) = \frac{1}{\mathcal{V}} \bigg(\sum_{p}  &  A_{pp}^2B_{pp}^2   + \sum_{p,q\ne p}\big( A_{pp}B_{pq}A_{qq}B_{qp} \nonumber \\
       + &  A_{pq}B_{qq}A_{qp}B_{pp}\big)\bigg).
\end{align}
This implies that if both $A$ and $B$ have no diagonal terms in the energy eigenbasis, the infinite-time value of the $F^{AB}(\infty)$ will be $0$.
At the same time, we can also state that if at least one of the operators $A$ or $B$ has non-zero diagonal elements, then $F^{AB}(\infty)$ can be different from zero.

\section{Emergence of algebraic relaxation}\label{sec:mechanism}

In this section we aim to first show that the combined effect of local Hamiltonian dynamics and algebraic scaling of the finite-size infinite-time OTOC, i.e. $F^{AB}(\infty)\sim 1/L^\alpha$ (where $L$ is the system size and $\alpha$ is a real positive number), results in the slow relaxation of the OTOC in time, i.e. $F^{AB}(t)\sim 1/t^\alpha$, Fig.\ref{fig:FigSummary}(a,b,d).
For local and bounded Hamiltonians, the speed of propagation of correlations is limited by the Lieb-Robinson bound \cite{LiebRobinson1972, CheneauKuhr2012}. This implies that at any given time, an accurate description of the OTOC of a system which is large (even in the thermodynamic limit), can be obtained simply considering a finite portion of it.  Given the Lieb-Robinson velocity $\vlr$, one can thus choose to study the evolution of the OTOC till time $t$, with a system size $L=s\;\vlr \;t$, where $s$ is a real number larger than $1$.
One then gets
\begin{align}
F^{AB}_{L=\infty}(t) \approx F^{AB}_{L=s\;\vlr \;t}(t) \label{eq:inf_to_finite_size}
\end{align}
where with the sub-index $L$ in $F^{AB}_{L}(t)$ we stress that the OTOC is computed for a system of size $L$. The main step is now to consider, for each system size $L$ (which itself is a function of time), the infinite
time value of the OTOC $F^{AB}_{L=s\;\vlr \;t}(\infty)$.
Even considering the case in which a system scrambles as fast as possible within the region of size $L$, the decay of $F^{AB}_{L=\infty}(t)$ can, at the fastest, be given by
\begin{align}
F^{AB}_{L=\infty}(t) \approx F^{AB}_{L=s\;\vlr \;t}(\infty) \label{eq:inf_time}
\end{align}
which, since $L$ increases with time, is a time-dependent quantity.
At this point, if one is able to show that $F^{AB}_{L}(\infty)$ decays algebraically with the system size, e.g. $F^{AB}_{L}(\infty)\propto L^{-\alpha}$, then the OTOC of the thermodynamic size system also decays algebraically, or more precisely from Eq.(\ref{eq:inf_time}) one can write
\begin{align}
F^{AB}_{L=\infty}(t) \propto \frac{1}{t^\alpha} \label{eq:slow_decay}
\end{align}
because $L=s\;\vlr \;t$.
The actual decay of the OTOC may even be slower, for example considering cases in which the system goes through prethermalization \cite{LuitzKhemani} or in which the system is many-body localized \cite{Lee2019}. However, the relaxation cannot be faster, hence the OTOC will have a slow relaxation, either algebraic or slower, see Fig.\ref{fig:FigSummary}(b).

For many-body Hamiltonians with power-law interactions it has been found that a light-cone for the propagation of the correlations may not exist but, instead, correlations may propagate as $t^\beta$ where $\beta$ is a real number larger than $1$ \cite{Foss-Feig2015, LuitzBarLev2019, ColmenarezLuitz2020, Kuwahara2020, Else2020}. Even in this case, analogously  using Eq.(\ref{eq:inf_to_finite_size},\ref{eq:inf_time}), one can get to the conclusion that
\begin{align}
F^{AB}_{L=\infty}(t) \propto \frac{1}{t^{\alpha\beta}}, \label{eq:slow_decay_longrange}
\end{align}
which is still an algebraic decay. In order for the OTOC to decay exponentially, one would then need the finite-size infinite-time OTOC $F^{AB}_{L}(\infty)$ to decay exponentially with the system size.

At this point, it is important to stress that even in bounded, local, time-dependent systems there is a finite velocity of propagation of correlations. For example, for the random unitary circuits discussed in \cite{NahumHaah2017, NahumHaah2018, KeyserlingkSondhi2018, RakovszkyKeyserlingk2018, KhemaniHuse2018}, the spread of correlation is limited by one extra site at each period.
It is also possible to extend this to continuously driven or periodically driven systems. In fact, given a chosen level of accuracy required to describe the system, one can do a Trotter decomposition of the evolution operator with local gates. Since the Hamiltonian is bounded, the same level of accuracy can be maintained throughout the evolution by the same time step $dt$ in the evolution operator. This, thus, results in mapping the evolution to local gates which, similarly to the case of random unitary circuits just discussed, results in a limit speed of propagation.
As an interesting side remark, this would not be possible for unbounded local Hamiltonians, e.g. for non-number-conserving bosonic systems, which in fact may not have a Lieb-Robinson bound.
In the following, we will also show that in time-dependent systems it is possible that the infinite-time OTOC scales algebraically with the system size, and thus the two sufficient ingredients for the emergence of slow relaxation of the OTOC can be present also for time-dependent systems.

\section{Number-conserving systems} \label{sec:time_ind_num_cons}
\subsection{Estimate of the infinite-time value of an OTOC}\label{sec:estimate_num_cons}
For a local Hamiltonian, following ETH one can write the matrix elements of local operators $A$ in the energy eigenbasis as \cite{Srednicki}
\begin{align}\label{eq:ethob}
A_{pq} = A(\bar{E})\delta_{pq} + e^{-S(\bar{E})/2} f_A(\bar{E}, \Delta_{pq}) R_{pq}
\end{align}
where $\bar{E}=(E_q+E_p)/2$ and $\Delta_{pq} = E_q - E_p$ and $R_{pq}$ is a random matrix with zero mean and unit variance. $S(\bar{E})$ is the thermodynamic entropy  and  is  related to the Hilbert space dimension as  $ e^{-S(\bar{E})/2}\sim \mathcal{V}^{-\frac{1}{2}}$.

For extensive, time-independent, systems one can write that \cite{Srednicki_1998, HuangZhang2019}
\begin{align}\label{eq:ethobav}
 A(\bar{E})\approx \tr(A)  +O\left(L^{-1}\right).
\end{align}
For number-conserving systems, only number-conserving operators $A$ can have a non-zero trace $\tr(A)$ because all solely non-number-conserving operators will connect eigenstates belonging to different number sectors. Considering for instance the operator $\sigma^z_l$, one realizes that
\begin{align}
\tr(\sigma^z_l)=\frac{L-2n}{L}   \label{eq:sz_numberconserving}
\end{align}
where $n$ is the number of spins up. So when $A=\sigma^z_l$ and $B=\sigma^z_{m\ne l}$, then one can readily evaluate $F^{AB}_L(\infty)$ in Eq.(\ref{eq:otocdiag}) as
\begin{align}
 F^{AB}_{L}(\infty) = & \frac{1}{\mathcal{V}}  \sum_{p} A_{pp}^2B_{pp}^2  + \frac{1}{\mathcal{V}}  \sum_{p,q\ne p} A_{pp}A_{qq}|B_{pq}|^2 \nonumber \\
  & + \frac{1}{\mathcal{V}}  \sum_{p,q\ne p} B_{pp}B_{qq}|A_{pq}|^2 \nonumber \\
  \approx   & \tr(\sigma^z_l)^{4} + \frac{2}{\mathcal{V}}\sum_{ p} A_{pp}A_{pp}\left[(B B^{\dagger})_{pp}-  B_{pp}^2\right]\nonumber \\
  \approx   & \tr(\sigma^z_l)^{4} + \frac{2 }{\mathcal{V}} \sum_{p} \left[ \tr{(B B^{\dagger})}- B_{pp}^2\right] \tr(\sigma^z_l)^{2}  \nonumber \\
    \approx   & \tr(\sigma^z_l)^{4} + 2  \tr(\sigma^z_l)^{2}(1-\tr(\sigma^z_l)^{2}) \nonumber \\
        \approx   & \left(\frac{L-2n}{L}\right)^{2}\left[2-\left(\frac{L-2n}{L}\right)^{2}\right].  \label{eq:otocdiagof}
\end{align}
This is pictorially described in Fig.\ref{fig:FigSummary}(c,d).
When, $A=\sigma^z_l$ and $B=\sigma^x_{m\ne l}$, then $\tr(\sigma^x_{m\ne l})=0$ because the system is number-conserving and the operator is not. It results that the first and third terms in Eq.(\ref{eq:otocdiag}) are zero and, since $\tr(B^\dagger B)=1$, we get
\begin{align}
 F^{AB}_{L}(\infty) = & \frac{1}{\mathcal{V}}  \sum_{p,q\ne p} A_{pp}A_{qq}|B_{pq}|^2 \nonumber \\
  \approx   & \tr(\sigma^z_l)^{2} = \left(\frac{L-2n}{L}\right)^{2}\label{eq:otocdiagofZX}
\end{align}

When both  $A$ and $B$ are not number-conserving, the diagonal elements $A_{pp}$ and $B_{pp}$ are zero in the energy eigenbasis which gives $F^{AB}_{L}(\infty) =0$.

For a periodically driven systems, Floquet ETH \cite{fritzsch2021}, results in
\begin{align}\label{eq:ethobft}
A(\bar{E}) \approx \tr( A ) +O\left(\frac{1}{\mathcal{V}^{1/2}} \right).
\end{align}
As a result, the estimates are also those given in Eq. (\ref{eq:otocdiagof})- (\ref{eq:otocdiagofZX}), with the only difference that the convergence will occur faster with the system size (there are no slow $1/L$ corrections in the convergence of $A(\bar{E})$ towards $\tr(A)$).

For thermalizing time-dependent but not periodic systems, e.g. random unitaries, one can also transfer the results for periodically driven systems. In fact, if we assume that a non-periodic time-dependent system thermalizes fast, e.g. approximately in a time $t$, then one can approximate the long-time evolution beyond time $t$ by repeating periodically the evolution dynamics that occurred before time $t$. As a result, the finite-size infinite-time value of the OTOC is well described by a periodically driven system which follows Floquet ETH.

\subsection{Emergence of algebraic relaxation} \label{sec:time_ind_num_cons_emergence}
In this section, we consider local non-integrable Hamiltonians which are number-conserving. Without loss of generality, we will consider spin-$1/2$ systems.
As shown in Sec.\ref{sec:mechanism} two ingredients which guarantee slow decay of the OTOC are the presence of a Lieb-Robinson bound (or a power-law propagation of correlations) and an algebraic dependence of the infinite-time OTOC with the system size. Since we consider systems which are local and bounded, a Lieb-Robinson bound exists.
What we need to prove is that the infinite-time OTOC scales algebraically
with the system size. Let us for simplicity consider a thermodynamically large system in the zero magnetization sector (our results can be readily
extended to other sectors). If we consider a portion of size $L$ of such infinite system at zero magnetization, and we analyze the possible configurations, we realize that in such a portion there are many different possible magnetization sectors appearing. In particular, the probability of finding a configuration with $n$ spins up and $L-n$ spins down, $p_{n,L}$,
is given by
\begin{align}
p_{n,L} = \mathcal{C}^{L}_n/2^{L} \label{eq:prob_disc}
\end{align}
where $\mathcal{C}_n^N = N!/[n!(N-n)!]$ is the binomial coefficient. For large enough $L$, the probability can be written in the rescaled variable $\kappa_n = n/L-1/2$ as
\begin{align}
p_{n,L} =\sqrt{\frac{2 L}{\pi} } \exp\left[ -2L\kappa_n^2 \right]. \label{eq:prob_disc_kappa}
\end{align}

For time-independent Hamiltonians, as it is exemplified in Fig.\ref{fig:Fig0a}(a), which will be discussed in detail in Sec.\ref{sec:tdnc}, each magnetization sector converges towards its own infinite-time value $F^{AB}_{L,n}(\infty)$ where the sub-indices $L$ and $n$ indicate respectively the system size and the number of spins up in system. This occurs as long as at least one of the two operators $A$ and $B$ has nonzero diagonal elements in the Hamiltonian eigenbasis. Extrapolated (dotted) lines to $1/\mathcal{V}\rightarrow0$ are the estimated infinite-time value of the OTOC in Eq.(\ref{eq:otocdiagof}). In Fig. \ref{fig:Fig4} and \ref{fig:Fig6}, we demonstrate the same aspect, i.e. that each different magnetization sector converges to the value we predict in Eq.(\ref{eq:otocdiagof}), for time-dependent systems.
As the system size increases we can write $F^{AB}_{L,n}(\infty)$ as a Taylor expansion
\begin{align}
F^{AB}_{L,n}(\infty) & = \sum_{m=0}^{\infty} \mathcal{F}_m \kappa_n^m \label{eq:Taylor_Expansion}
\end{align}
where $\mathcal{F}_m$ are real coefficients.
We can thus write
\begin{align}
F^{AB}_{L}(\infty) &= \sum_n F^{AB}_{L,n}(\infty) p_{n,L} \nonumber \\
&\approx \int_{-\infty}^{\infty} d\kappa_n \left[ \sum_{m=0}^{\infty} \mathcal{F}_m \kappa_n^m \right] \sqrt{\frac{2 L}{\pi} } \exp\left[ -2L\kappa_n^2 \right] \nonumber \\
& \approx \mathcal{F}_0 + \frac{\mathcal{F}_2}{L} + O\left(\frac{ 1}{ L^2}\right). \label{eq:scaling_inifite}
\end{align}
This implies an algebraic decay of the OTOC with the system size which then results, following Eq.(\ref{eq:slow_decay}), in a slow decay of the OTOC.

\subsection{Example of time-independent Hamiltonian with local conserved quantities}\label{sec:tdnc}
For concreteness we study an $XXZ$ chain plus next-nearest neighbor terms
\begin{align}
H_{NNN} & = \sum_{l=1}^{L-1}\left[J\left( \sx_l\sx_{l+1} + \sy_l\sy_{l+1} \right) + J^z \sz_l\sz_{l+1}\right]  \nonumber \\
        & + \alpha\sum_{l=1}^{L-2} \left[J\left( \sx_l\sx_{l+2} + \sy_l\sy_{l+2} \right) + J^z \sz_l\sz_{l+2} \right]       \label{eq:XXZ_NNN}
\end{align}
where $J$ is a tunneling amplitude, $J^z/J$ the anisotropy and $\alpha$ is the relative weight of the next-nearest-neighbor terms. For early works studying thermalization with this Hamiltonian see \cite{Rigol2009, SantosRigol2010, Borgonovi2019a, Schiulaz2019}.
For $\alpha=0$ the system is integrable, while $\alpha\ne 0$ breaks the integrability. This can be shown by studying the level spacing statistics which typically follows a Poisson distribution for integrable systems and a Wigner-Dyson distribution for non-integrable ones \cite{BGS1984, Casati1980}. In particular, considering $\delta_n= E_{n+1}-E_n$ the level spacing between two consecutive energy levels $E_n$ and $E_{n+1}$
within a single symmetry sector, one can define the ratio $r_n = \max(\delta_n,\delta_{n+1})/\min(\delta_n,\delta_{n+1})$ and take an average $r=\sum_n r_n/N$ where $N$ is the number of energy level differences considered. For a Poisson distribution, $r$ can be computed analytically and it gives $r=2\ln2-1\approx 0.386$, while for a Wigner-Dyson distribution $r$ can be evaluated numerically to be $r\approx 0.529$ \cite{OganesyanHuse2007}.
In the current work, we use parameters $J^z/J=2$ and $\alpha=1$ which result in $r\approx 0.53$ already for a system size of $14$ spins.

\begin{figure}[h]
\includegraphics[width=\columnwidth]{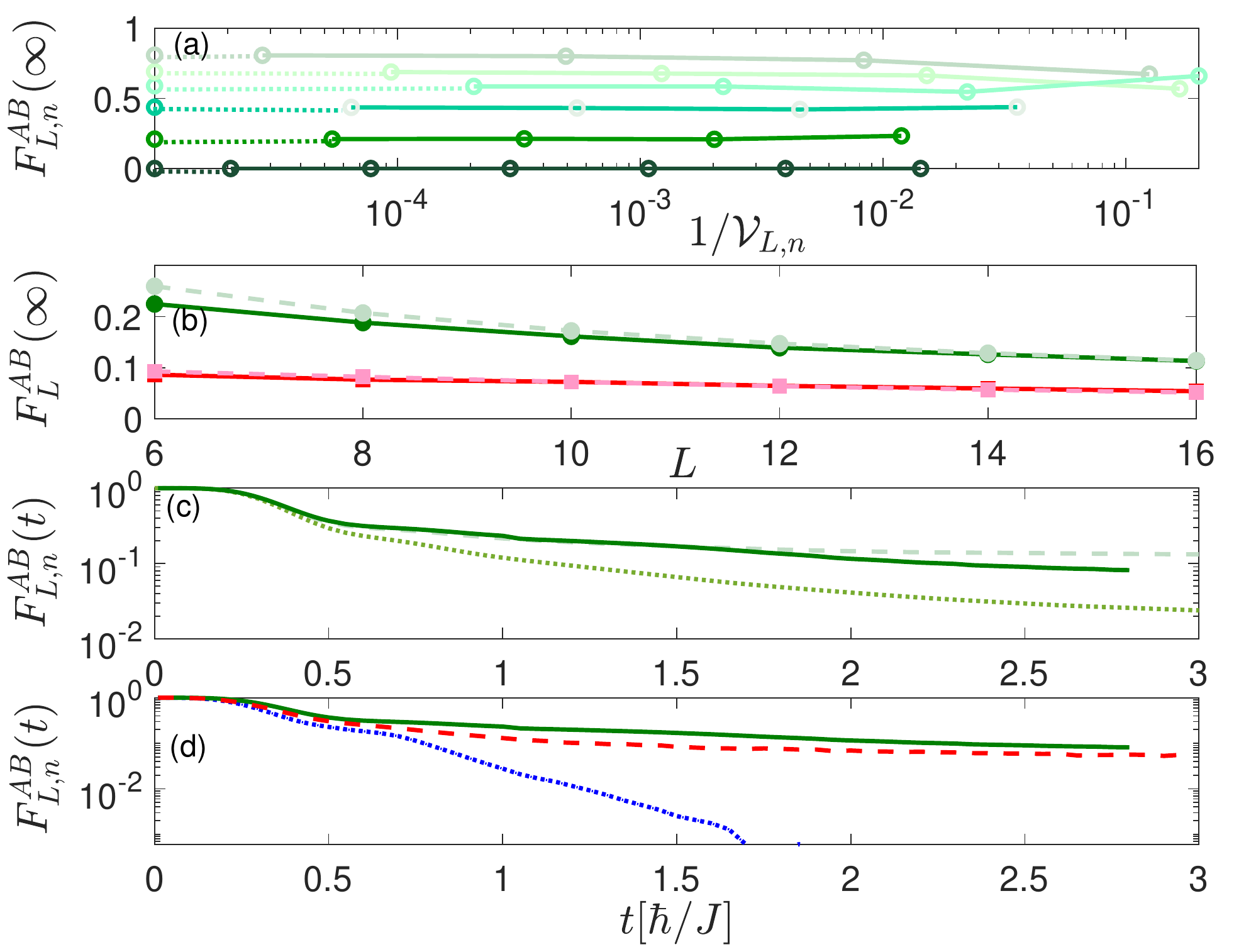}
\caption{(a) Infinite-time values of the OTOC $F^{AB}_{L,n}(\infty)$ for $A=\sigma^z_{L/2-1}$ and $B=\sigma^z_{L/2+2}$ in different magnetization subspaces of a non-integrable XXZ chain with Hilbert space size $\V_{L,n}$. The curves from top to bottom correspond to $n/L$ equal to $1/8$, $1/6$, $1/5$, $1/4$, $1/3$ and $1/2$ respectively. Solid lines are the values from exact diagonalization calculations, whereas dotted lines show the extrapolation to the estimations in Eq.(\ref{eq:otocdiagof}).
(b) Finite-size infinite-time values of the OTOC $F^{AB}_{L}(\infty)$ of the corresponding XXZ chain. Green solid line with \small{{$\circ$}} is for $A=\sz_{L/2-1}$ and $B=\sz_{L/2+2}$ and red solid line with \small{{$\square$}} for $A=\sz_{L/2-1}$ and $B=\sx_{L/2+2}$. The lighter dashed lines correspond to estimates from Eqs.(\ref{eq:otocdiagof}) and (\ref{eq:otocdiagofZX}) for the curves with the same color and symbol.
(c) Time evolution of the OTOC $F^{AB}_{L,n}(t)$ for the corresponding chain in the zero magnetization subspace where $n=L/2$ with $A=\sz_{L/2-1}$ and $B=\sz_{L/2+2}$. The green solid line is for a chain length $L=64$ using MPS evolution, and dotted green line is for a chain with $L=16$ in the zero magnetization sector only and using exact diagonalization. The dashed line is for the evolution of an infinite chain with the exact diagonalization results of chains of $L=16$ but considering all the magnetization sectors and using Eq.(\ref{eq:time_dep_scaling}).
(d) Time evolution of the OTOC $F^{AB}_{L,n}(t)$ in zero magnetization subspace for the corresponding chain with $L=64$. MPS evolutions are averaged over $400$ initial conditions. Green solid line is for $A=\sz_{L/2-1}$ and $B=\sz_{L/2+2}$, blue dotted line  for $A=\sx_{L/2-1}$ and $B=\sx_{L/2+2}$ and red dashed line is for $A=\sz_{L/2-1}$ and $B=\sx_{L/2+2}$.
In all panels, the Hamiltonian parameters are $J^z/J=2$ and $\alpha=1$. }
\label{fig:Fig0a}
\end{figure}

In order to test the validity of the various steps taken in Secs.\ref{sec:mechanism} and \ref{sec:time_ind_num_cons} and their ensuing conclusions, we have run numerical computations with the Hamiltonian (\ref{eq:XXZ_NNN}). In Fig.\ref{fig:Fig0a}(a) we show the value of $F^{AB}_{L,n}(\infty)$ as a function of the Hilbert space dimension $\V_{L,n}$ for different number sectors, which is a function of the number of spins up
$n$ and size of the system $L$. From top to bottom we show results
for $n/L=1/8$, $1/6$, $1/5$, $1/4$, $1/3$ and $1/2$ for the operators $A=\sigma^z_{L/2-1}$ and $B=\sigma^z_{L/2+2}$. These calculations were
done for  sizes up to $\V_{L,n}=36000$ with exact diagonalization. For $n/L=1/2$, because of particle-hole symmetry, all the diagonal elements
in the energy eigenbasis are $0$, and hence the infinite-time value of $F^{AB}_{L,n}(\infty)$ is $0$ \footnote{For particle-hole symmetry, $\Ss\sz\Ss^{-1}=-\sz$ and hence, $\Ss\sz=-\sz\Ss$ which implies $\langle n|\Ss\sz| n\rangle=-\langle n |\sz\Ss| n\rangle$. As, $[H,\Ss]=0$, and $\Ss^2|n\rangle=|n\rangle$, one gets that $\langle n|\sz| n\rangle=-\langle n|\sz | n\rangle$ which is true only if $\langle n|\sz| n\rangle=0$.}.
Fig.\ref{fig:Fig0a}(a) shows that
each magnetization sector is converging towards a different value given by Eq.(\ref{eq:otocdiagof}).

In Fig.\ref{fig:Fig0a}(b), we show $F^{AB}_{L}(\infty)$ for $A=\sigma^z_{L/2-1}$ and $B=\sigma^z_{L/2+2}$ in solid lines with $\circ$, and for $A=\sigma^z_{L/2-1}$ and $B=\sigma^x_{L/2+2}$ with a solid line with $\square$.
The dashed lines correspond to estimates from the first line of Eq.(\ref{eq:scaling_inifite}), computed for the cases with the same color and symbols used for the solid lines. As the system size increases, we find a remarkable match between the solid and their corresponding dashed lines, and in both cases we observe that the infinite-time OTOC $F^{AB}_{L}(\infty)$ is decaying slowly with the system size.
One may wonder why, when studying the infinite-time OTOC of a thermodynamically large systems in the $0$ magnetization sector, we consider all the possible configurations for a system of size $L$. In fact, if one was considering only the $0$ magnetization sector for each size $L$, then he/she would obtain always $0$ for the infinite-time OTOC. However, as stressed earlier, in a finite size $L$ of a thermodynamically large system in the $0$ magnetization sector all possible configurations are equally probable, and using all of these configurations gives a much more accurate description of the system dynamics.
This is confirmed in Fig.\ref{fig:Fig0a}(c) where we show the time evolution of $F^{AB}(t)$ in the zero magnetization subspace for $A=\sigma^z_{L/2-1}$ and $B=\sigma^z_{L/2+2}$ using green lines. Dotted green lines are computed using a chain of length $L=16$ in the zero magnetization sector ($n=8$) using exact diagonalization, whereas the solid lines are for $L=64$ using MPS evolution with $400$ different states in the zero magnetization sector sampled from a uniform distribution. The dashed line in Fig.\ref{fig:Fig0a}(c) shows the value of $F^{AB}(t)$ computed from
\begin{align}
F^{AB}(t)\approx \sum_n F^{AB}_{L,n}(t)p_{n,L} \label{eq:time_dep_scaling}
\end{align}
confirming that the OTOC computed for a certain system size can be well approximated by the weighted average of the ones computed in different number sectors on a smaller system size. In particular, we have used $L=16$, for which we can readily use exact diagonalization. The dashed curve shows that for times $t \lesssim 2$, the choice of $L=16$ is large enough to predict accurately the evolution for a system of size $L=64$ in the zero magnetization sector ($n=32$) at short times.
Instead, the dotted line, which considers only the zero magnetization sector of a chain of $L=16$ spins, shows already significant difference from the large $L=64$ at times $t=0.5$.

As we have anticipated, the choice of the operators inside the OTOC is very important. This is clear from Fig.\ref{fig:Fig0a}(d) where the green solid lines are for $A=\sigma^z_{L/2-1}$ and $B=\sigma^z_{L/2+2}$, the red dashed lines for $A=\sigma^z_{L/2-1}$ and $B=\sigma^x_{L/2+2}$, and the blue dotted lines for $A=\sigma^x_{L/2-1}$ and $B=\sigma^x_{L/2+2}$. In the first case, both operators are number-conserving and thus they have nonzero diagonal elements in the energy eigenbasis. Hence, in this case, one can readily apply the derivation in Sec.\ref{sec:mechanism} and expect an algebraic relaxation. In the second case, one operator is non-number-conserving, thus resulting in $0$ diagonal elements. However, one can see from Eq.(\ref{eq:otocdiag}) that the infinite-time value of $F^{AB}_{L,n}$ is typically nonzero and can be different in different number sectors. Once again, one can apply the theory in Sec.\ref{sec:mechanism} and expect slow relaxation. In the third case, both operators are non-number-conserving and there is thus no difference between the possible infinite-time values of the OTOC in different number sectors, i.e. in Eq.(\ref{eq:scaling_inifite}) one would have $F^{AB}_{L,n}(\infty)=0$. There is thus no constraint that forces the OTOC dynamics to be algebraic, and one observes fast decay.

\subsection{Examples of time-dependent Hamiltonians with local conserved quantities}\label{sec:tdnnc}

In the case of periodic local and number-conserving Hamiltonians, the effective Hamiltonian $\He$ from Eq.(\ref{eq:He}) will take central stage.
In general, $\He$ can have very different properties from the instantaneous $H(t)$, for instance $H(t)$ can instantaneously be integrable and local, while $\He$ could be non-integrable and nonlocal. However if $H(t)$ is number-conserving at each point in time, so is $\He$.
Furthermore, as discussed previously, the locality of the Hamiltonian used in the dynamics guarantees the existence of a speed limit for the propagation of the correlations.
Note that while $\He$ is not uniquely defined, as one can always add or remove $2\pi/T$ to it, its eigenvectors are, and hence the value of the infinite-time OTOC in Eq.(\ref{eq:otocdiag}) is unique.
Hence, the previous analysis which led to Eq.(\ref{eq:slow_decay}) can be
readily transferred to periodically driven systems when considering the effective Hamiltonian $\He$. Each number sector can decay to a different asymptotic value, hence the dependence of the infinite-time OTOC with the
system size will follow an algebraic decay. This will result in a slow decay of the OTOC.
In fact, it follows that stroboscopically the system relaxes slowly which
implies that, overall, the OTOC goes through a slow decay for the time-dependent $H(t)$.

\begin{figure}
\includegraphics[width=\columnwidth]{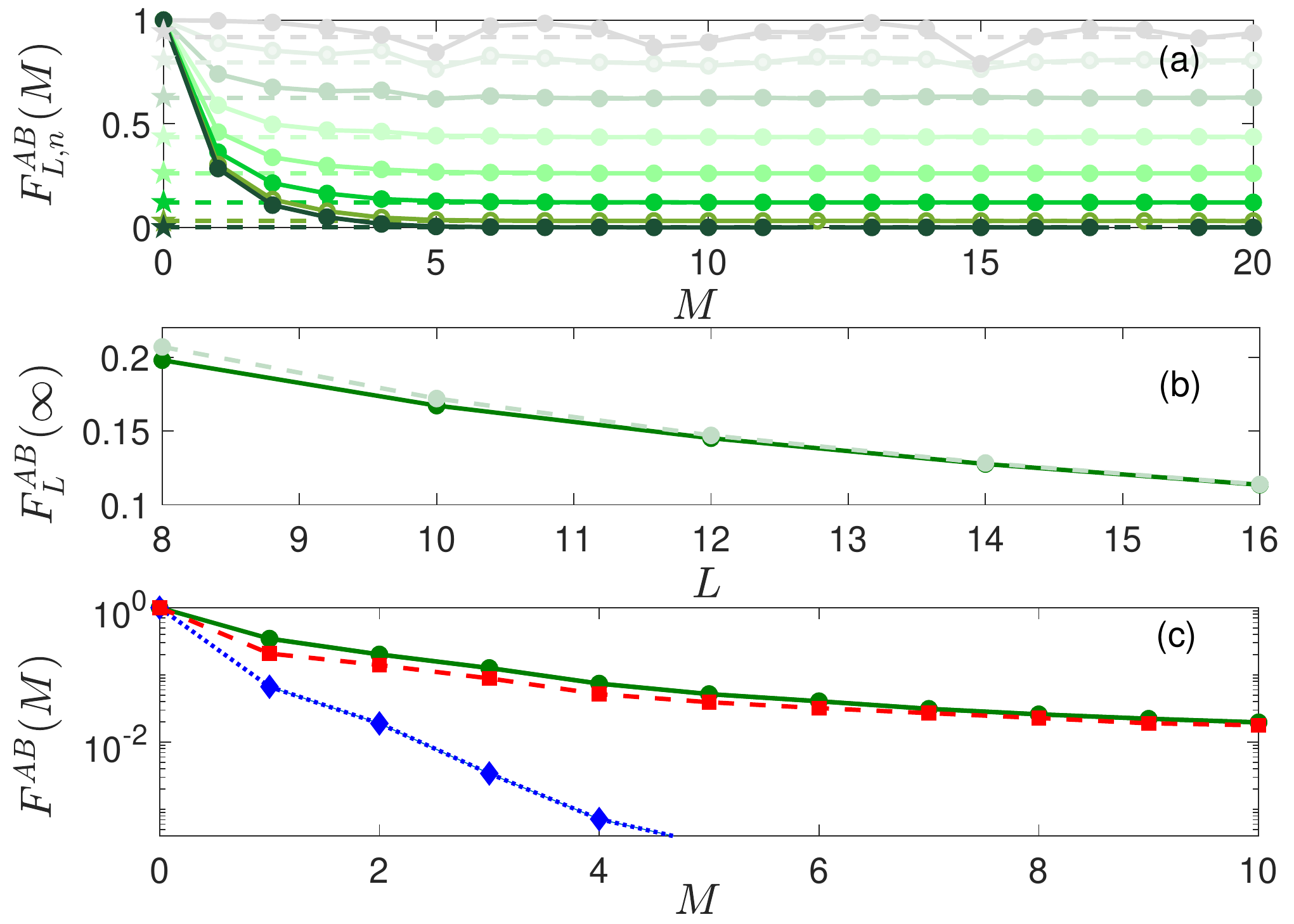}
\caption{(a) Time evolution of the OTOC $F^{AB}_{L,n}(M)$, after $M$ periods, in different magnetization subspace of a time-dependent non-integrable XXZ chain of length $L=16$ given by Eq.(\ref{eq:kicked_Ham}). The curves from top to bottom correspond to the filling $n/L$ equal to $1/16$, $2/16$, $3/16$, $4/16$, $5/16$, $6/16$, $7/16$ and $8/16$ respectively. Pentagram depicted at $M=0$ indicates the infinite-time estimates discussed in Eq.(\ref{eq:otocdiagof}). Dashed lines correspond to the values obtained by using effective Hamiltonian $\He$ for evolution. (b) Finite-size infinite-time values of the OTOC $F^{AB}_{L}(\infty)$ (dark green line) of the corresponding  XXZ chain. The dashed green line denotes the estimates provided from the first line of Eq.(\ref{eq:scaling_inifite}) and Eq.(\ref{eq:otocdiagof}).
For both panels (a) and (b), we have chosen $A=\sz_{L/2}$ and $B=\sz_{L/2+1}$.
(c) Time evolution of the OTOC $F^{AB}(M)$ of the corresponding non-integrable XXZ chain of length $L=30$. The green solid line with $\circ$ is for $A=\sz_{L/2}$ and $B=\sz_{L/2+1}$, the blue dotted line with $\diamond$ for $A=\sx_{L/2}$ and $B=\sx_{L/2+1}$ and the red dashed line with $\square$ for $A=\sz_{L/2}$ and $B=\sx_{L/2+1}$. Results are averaged over $100$ initial conditions.}
\label{fig:Fig4}
\end{figure}

As an illustrative example, we consider the following kicked model, similar to the one studied in \cite{RakovszkyKeyserlingk2018} where two separate Hamiltonians $H_{k1}$ and $H_{k2}$ are switched at each half period. The time evolution operator for one period $T$ takes the form
\begin{align}
U_T& = e^{-\im  \frac{T}{2}H_{k1}} e^{-\im  \frac{ T}{2}H_{k2}},
\end{align}
with
\begin{align}
H_{k1} & =\sum_l[ J\left( \sx_l\sx_{l+1} + \sy_l\sy_{l+1} \right) + J^z
\sz_l\sz_{l+1}],  \nonumber \\
H_{k2} & = \sum_l[J\left( \sx_l\sx_{l+1} + \sy_l\sy_{l+1} \right) +   J^{z2} \sz_l\sz_{l+2}] ,       \label{eq:kicked_Ham}
\end{align}
Here, we use $J=(2\sqrt{3}+3)/7$, $J^z=(\sqrt{3})+5)/6$ and $J^{z2}=\sqrt{5}/2$. The average level spacing ratio $r\approx0.53$ for $L>10$.
For different number sectors the OTOC converges to different values. In Fig.\ref{fig:Fig4}(a) we have considered a system with $L=16$ and evolved in time the OTOC $F^{AB}_{L,n}$ for different number sectors, and from top to bottom we consider fillings from $1/16$ to $1/2$ which are shown by the solid lines. The dashed lines show the corresponding infinite-time value computed using the effective Hamiltonian $\He$. Crucially, they coincide with the estimates in Eq.(\ref{eq:otocdiagof})
which are denoted by pentagrams on the $y-$axis.
In Fig.\ref{fig:Fig4}(b) we show the slow relaxation of the infinite-time value of the OTOC $F^{AB}_L(\infty)$ versus system size, considering $A=\sigma^z_{L/2}$ and $B=\sigma^z_{L/2+1}$ for the kicked Hamiltonian Eq.(\ref{eq:kicked_Ham}) with period $T=1$. It is clear from the figure that the numerically exact values (solid dark green line) are in close agreement with the estimates (dashed light green line) from Eq.(\ref{eq:otocdiagof}) and the first line of Eq.(\ref{eq:scaling_inifite}).
And as expected, in Fig.\ref{fig:Fig4}(c) we show that only the relaxation dynamics for the case in which both operators are not number-conserving is fast, the OTOC with $A=\sigma^x_{L/2}$ and $B=\sigma^x_{L/2+1}$ depicted by the blue dotted line, while the cases in which at least one operator is number-conserving, e.g. the OTOC with $A=\sigma^z_{L/2}$ and $B=\sigma^x_{L/2+1}$ in dashed red line and OTOC with $A=\sigma^z_{L/2}$ and $B=\sigma^z_{L/2+1}$ in solid green line, are slow.

In the example above we considered a periodically driven system, for which there always exists an effective Hamiltonian $\He$. The same cannot be told for aperiodic driven systems as, for instance, the use of random unitaries \cite{RakovszkyKeyserlingk2018, KhemaniHuse2018}.
However, as discussed in Sec.\ref{sec:mechanism}, for emergence of slow relaxation of the OTOC, it is sufficient to have an algebraic propagation of correlations (e.g. the Lieb-Robinson bound), and a system-size dependence of the infinite-time OTOC. Even considering nearest neighbor two-spins number-conserving random unitaries one has all the ingredients for the emergence of a slow decay of the OTOC. Since the random unitaries only act on nearest neighbour spins, there still exist a bound on the speed of the spread of correlations. Furthermore, for different number sectors the infinite-time OTOC is different.
It is thus possible to apply the same arguments as above, Eqs.(\ref{eq:otocdiagof},\ref{eq:otocdiagofZX}) and Eq.(\ref{eq:slow_decay}), to predict a slow relaxation of the OTOC.
However in this case we will have to consider $\He_t$ from Eq.(\ref{eq:Het}).
As discussed for periodically driven systems, if $H(t)$ is number-conserving at each point in time, then $\He_t$ is also number-conserving. The key point of our analysis is this: if the evolution up to time $t$ can be described by the evolution of a {\it time-independent} Hamiltonian which would lead to slow relaxation of the OTOC, then the relaxation has to be slow up to that time. And if this can be done for an arbitrary time, then the relaxation is always slow.

To illustrate this idea, we consider a constrained random unitary circuit as in \cite{RakovszkyKeyserlingk2018, KhemaniHuse2018}. More precisely we consider a number-conserving circuit of repeated and different random evolutions, each consisting of two layers of two site unitary gates acting on even and odd bonds at even and odd times respectively, and described by the evolution operator $U(M)=\prod_{m=1}^M U(m)$ where

\begin{equation}
  U(m)= \left(\prod_{l} U_{2l,2l+1} \right)\left( \prod_{l}  U_{2l-1,2l} \right) \label{eq:circuit_Ham}
\end{equation}

Because of number conservation, each two-site gate is block diagonal with respect to $k$, the total number of spins ups (or downs) on the two sites. Each of these blocks is Haar-random and chosen independent of the
other. The structure of the two site gate $U(m)$ is of the form
\begin{equation}
U(m)=
\begin{pmatrix}
  U_{k=0}(m) &  &  \\
   & U_{k=1}(m) &  \\
   &  & U_{k=2}(m)
\end{pmatrix},
\end{equation}
where $U_{k=0}(m)$ and $U_{k=2}(m)$ blocks are of size $1\times1$ (a scalar) and $U_{k=1}(m)$ block is $2\times2$ Haar-random unitary.
\begin{figure}
\includegraphics[width=\columnwidth]{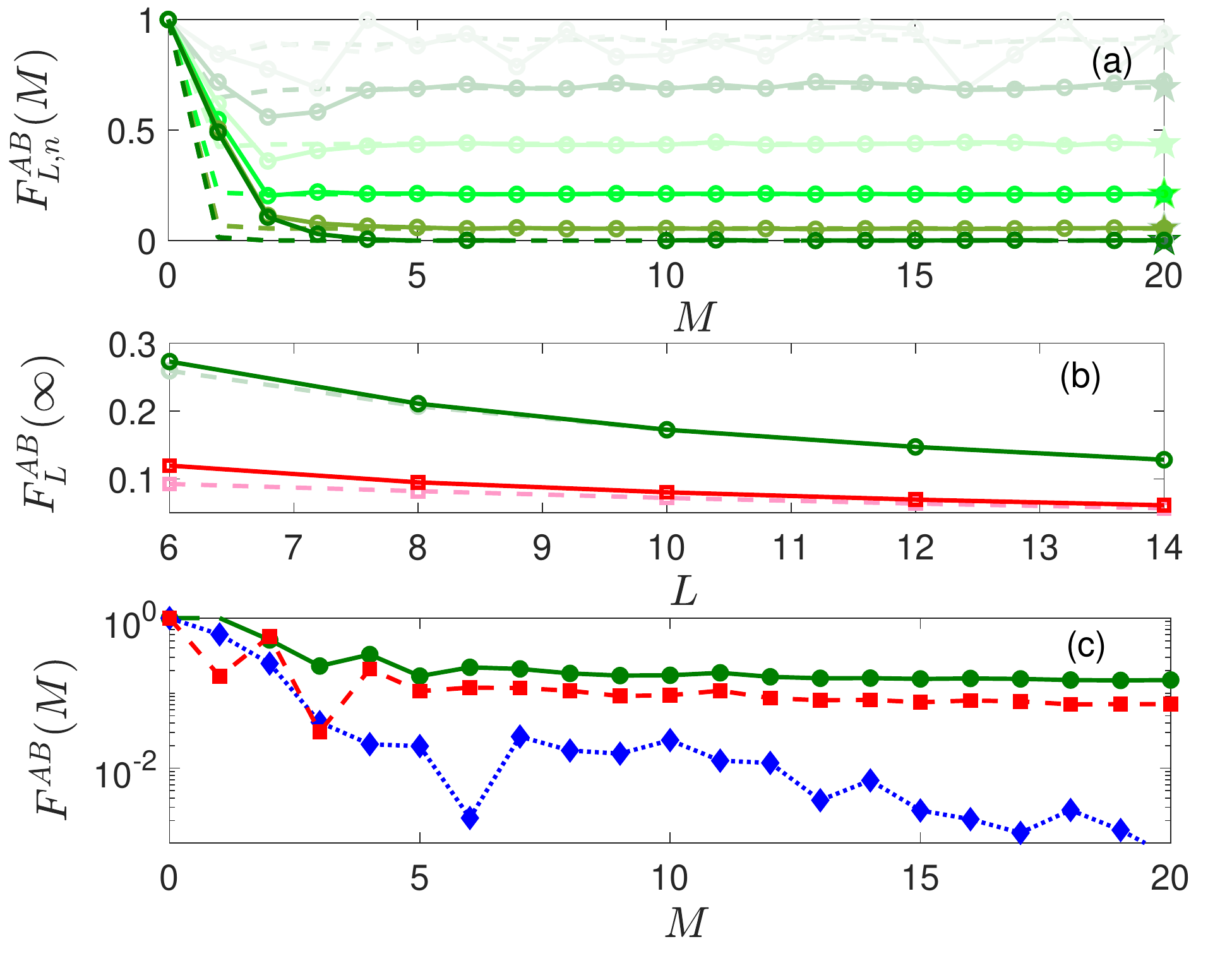}
\caption{(a) Time evolution of  OTOC $F^{AB}_{L,n}(M)$ in different magnetization subspaces for a constrained random unitary circuit of length $L=12$. The curves from top to bottom correspond to the filling $n/L$ equal to $1/12$, $2/12$, $3/12$, $4/12$, $5/12$ and $6/12$ respectively. Estimates for each filling from Eq.(\ref{eq:otocdiagof}) are shown by pentagrams at $M=20$. Dashed lines correspond to the values of OTOC obtained from the effective Hamiltonian $\He_t$.  Here, $A=\sz_{L/2}$ and $B=\sz_{L/2+1}$. (b) Finite-size infinite-time values of the OTOC $F^{AB}_{L}(\infty)$ of a constrained random unitary circuit. The green solid line with $\circ$ corresponds to $A=\sz_{L/2}$ and $B=\sz_{L/2+1}$ and the red solid line to $A=\sz_{L/2}$ and $B=\sx_{L/2+1}$. The lighter dashed curves correspond to estimates from the first line of Eq.(\ref{eq:scaling_inifite}) and Eqs.(\ref{eq:otocdiagof}) or (\ref{eq:otocdiagofZX}) respectively for the cases with the same color and symbol. (c) Time evolution of the OTOC $F^{AB}(M)$ of a constrained random unitary circuit of length $L=12$. The green solid line with $\circ$ is for $A=\sz_{L/2}$ and $B=\sz_{L/2+1}$, the blue dotted line with $\diamond$ for $A=\sx_{L/2}$ and $B=\sx_{L/2+1}$ and the red dashed line with $\square$ for $A=\sz_{L/2}$ and $B=\sx_{L/2+1}$. For all panels, the results are from a single random realization of the evolution. }
\label{fig:Fig6}
\end{figure}
Fig.\ref{fig:Fig6} mirrors Fig.\ref{fig:Fig4} in all of its panels.
Fig.\ref{fig:Fig6}(a) focuses on the evolution of the OTOC in each number sector. Again, we observe that within each sector the asymptotic value of $F^{AB}_{L,n}(\infty)$ stabilizes to different values for different quantum numbers $n$, and the predictions from Eq.(\ref{eq:otocdiagof}) are marked by a pentagram. Note that, here, at each time step $M$ we compute the effective Hamiltonian $\He_t$ and then we compute the instantaneous value of $F^{AB}_{L,n}(\infty)$ for each number sector $n$ which is represented by dashed lines.
In panel \ref{fig:Fig6}(b) we show the dependence of the OTOC versus the system size. Here, the values $F^{AB}_{L}(\infty)$  are obtained by calculating $\He_t$ at time $M=20$. Using the light dashed lines we show that an accurate description of the computed values are provided by Eqs.(\ref{eq:otocdiagof},\ref{eq:otocdiagofZX}) and the first line of Eq.(\ref{eq:scaling_inifite}).
Fig.\ref{fig:Fig6}(c) shows the relaxation of different OTOCs in time, which is slow when at least one of the two operators $A$, $B$ is number-conserving, green solid line with $\circ$ (where specifically $A=\sz_{L/2}$ and $B=\sz_{L/2+1}$) or red dashed line with $\square$ ($A=\sz_{L/2}$ and $B=\sx_{L/2+1}$), and fast when none of them is, blue dotted line with $\diamond$ ($A=\sx_{L/2}$ and $B=\sx_{L/2+1}$).
We stress here that all panels in Fig.\ref{fig:Fig6} are obtained by just a single realization of the random evolution, without any averaging between different realizations. In fact, the averaging would only result in smoother curves but does not fundamentally affect the results.

\section{Time-independent non-number-conserving Hamiltonians} \label{sec:time_ind_non_num_cons}
Our final step is to extend the results to non-number-conserving systems with just energy conservation. In the absence of any number conservation, the infinite-temperature average of a traceless local observable is obviously zero, i.e. $\langle A \rangle=0$.
However, following \cite{HuangZhang2019} one can write, using $H=\sum_l H_l$ where $H_l$ is a local component of the Hamiltonian,
\begin{align}
F^{AB}_{L}(\infty) &\approx \frac{1}{L} \frac{ \langle A A^\dagger \rangle |\langle HB \rangle|^2 + \langle B B^\dagger \rangle |\langle HA \rangle|^2 }{ \langle H H_l \rangle }  \nonumber \\
  &\approx \frac{1}{L} \frac{ |\langle HB \rangle|^2 + |\langle HA \rangle|^2 }{ \langle H_l^2 \rangle }.    \label{eq:inft}
\end{align}
Here we have used the fact that $A$ and $B$ are unitary, the locality of $H_l$ which implies that $\tr(H H_l)=\tr(H_l^2)$, and we have neglected terms that decay faster than $1/L$.
From Eqs.(\ref{eq:inf_to_finite_size}-\ref{eq:slow_decay}) one can thus predict a slow relaxation of the OTOC, as soon as $\tr(HA)\ne 0$ or $\tr(HB)\ne 0$.

\subsection{Examples} \label{sec:non_num_cons_examples}
\begin{figure}
\includegraphics[width=\columnwidth]{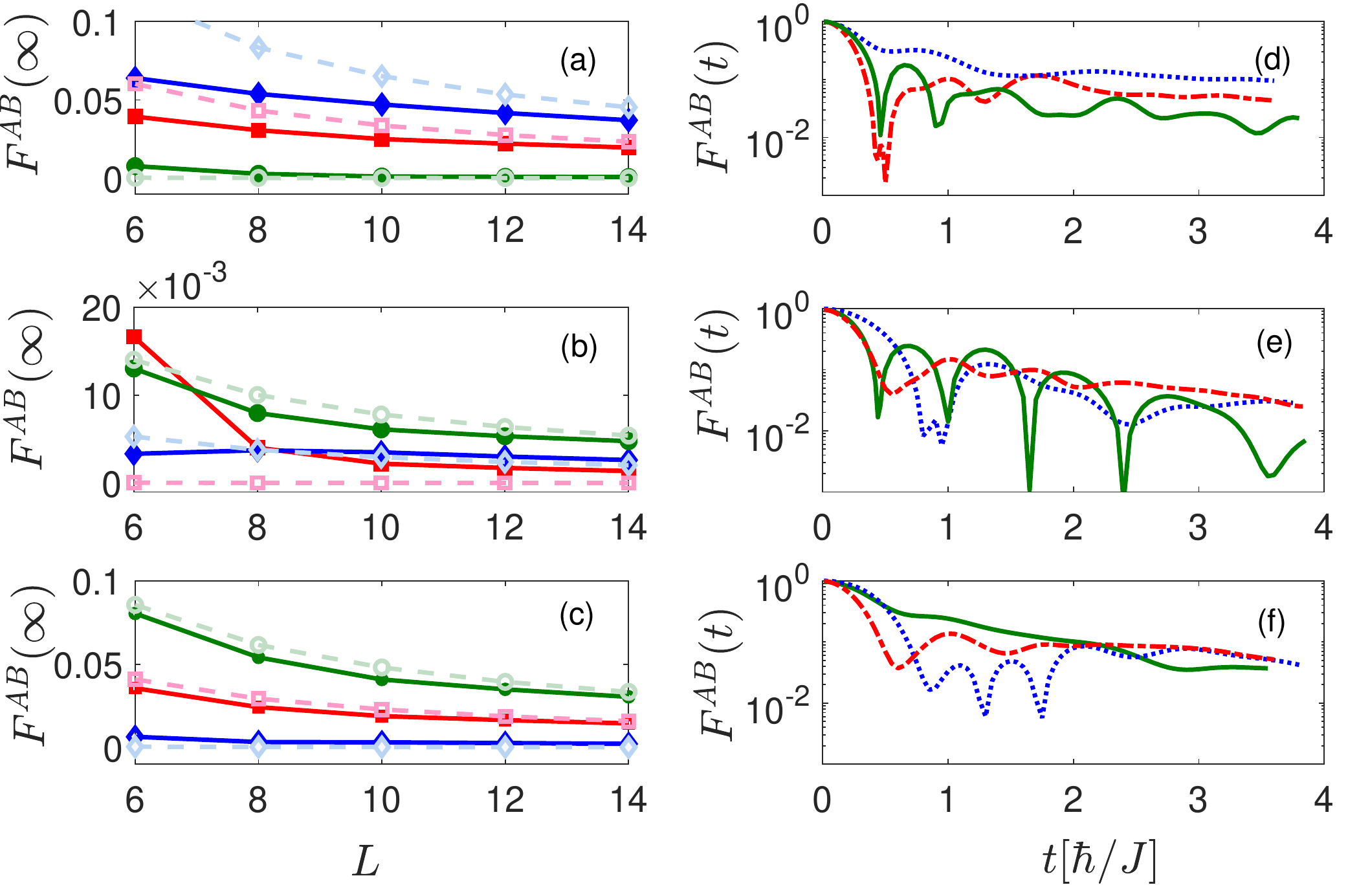}
\caption{Panels (a)-(c): Bright solid lines correspond to the scaling of the infinite-time value of the OTOC $F^{AB}_{L}(\infty)$ of an non-integrable typical Hamiltonian in Eq.(\ref{eq:typical}). Shaded colors correspond to the estimates from Eq.(\ref{eq:inft}).
Panels (d)-(f): Time evolution of the OTOC $F^{AB}(t)$ for the Hamiltonian in left panels for system size $L=30$. Other parameters: $A=\sz_{L/2}$, $B=\sz_{L/2+1}$ for panels (a) and (d), $A=\sx_{L/2}$, $B=\sx_{L/2+1}$ for panels (b) and (e), and $A=\sy_{L/2}$, $B=\sy_{L/2+1}$ for panels (c) and (f).  Different colors correspond to three random realizations of parameters $J_{nm}$ and $h_n$ in Eq.(\ref{eq:typical}), see \cite{fnt} for details, and the results for the time evolution are averaged over $100$ random initial conditions. In the left panels, realizations are differentiated with the symbols $\circ$  for green, $\square$ for red and $\diamond$ for blue line whereas in right panels, they are shown by solid lines for green, dotted lines for blue and dot dashed for red color. }
\label{fig:Fig1m}
\end{figure}

To give a glimpse of the generality of the result discussed here, we consider three different random time-independent local Hamiltonians described
by
\begin{align}
H_{tp} & = \sum_{l=1}^{L-1}  \sum_{n,m=x,y,z} J_{nm} \sigma^n_l\sigma^m_{l+1}   + \sum_{l=1}^{L} \sum_{n=x,y,z} h_n \sigma^n_l      \label{eq:typical}
\end{align}
Here, $J_{nm}$ and $h_n$ are random numbers uniformly distributed in the interval $[0,1]$. The model is non-integrable which we confirmed by calculating $r\approx 0.6$ for all $3$ different random realizations of Eq. \ref{eq:typical} we studied. The explicit parameters are given in \cite{fnt}.
The polynomial scaling of infinite-time OTOC is probed in the left panels of Fig. \ref{fig:Fig1m} for all the three realizations.   Panels (a) and (d) are for operators $A=\sz_{L/2}$ and $B=\sz_{L/2+1}$, (b) and (e) for $A=\sx_{L/2}$ and $B=\sx_{L/2+1}$ and (c) and (f) for $A=\sy_{L/2}$ and $B=\sy_{L/2+1}$. In these plots different Hamiltonians are depicted with curves of different colors and symbols (blue $\diamond$, red $\square$ and green $\circ$) or line style. The dashed lines in panels (a)-(c) correspond to the estimates given from Eq.(\ref{eq:inft}), while solid lines are used for exact numerical results.
With the increase of system size, the exact OTOC values converge towards the estimated ones.
In panels (d)-(f) we show the time evolution of the three different OTOCs versus time. In each of the (d)-(f) panels, the different curves correspond to different Hamiltonian. Specifically, the green solid, the red dashed and the blue dotted lines correspond respectively to the green line with $\circ$, the red line with $\square$ and the blue line with $\diamond$ in panels (a)-(c).
The infinite-time value of the OTOC is constrained by the magnitude of the overlap of the observables $A$ and $B$ with the Hamiltonian $H$, i.e. when $\tr(HA)$ and/or  $\tr(HB)$ is larger, then the value of $F^{AB}_{L}(\infty)$ is also larger. For instance, $\tr(H\sz_{L/2})\approx 0.0487$, $\tr(H\sx_{L/2})\approx 0.2891$ and $\tr(H\sy_{L/2})\approx 0.721$ for the realization indicated by green solide lines in panels (d)-(f) [and green solid lines with circles in (a)-(c)], which are larger in magnitude in panel (f), followed by panel (e) and (d) of Fig. \ref{fig:Fig1m}.
We emphasize that we use translationally invariant
Hamiltonians and hence the overlap $\tr(H A)$ or $\tr(H B)$ is independent of the system size. A small value of $\tr(H A)$ may result in a very low value of $F^{AB}_{L}(\infty)$. Because of this, at early times the OTOC seems to decay faster or even exponentially. However, this is a transient effect which vanishes at long enough times and, indeed, long time dynamics of larger systems are required to capture the algebraic decay under these circumstances. For this reason, in Fig.\ref{fig:Fig1m}(d-f) we have studied a system of size $L=30$.
We stress here that we have purposely shown the OTOC relaxation for the three random cases in Fig.\ref{fig:Fig1m} to highlight clearly that in typical, local, and time-independent systems, slow relaxation is generic.
\section{Conclusions}\label{sec:conclusions}
In the recent years, the OTOC has emerged as a diagnostic tool for scrambling.
Here we have considered deterministic many-body systems and we have shown conditions for the emergence of an algebraic relaxation of the OTOC.
We have found two sufficient conditions to provide, typically, algebraic decay of the OTOC:  $1)$ a Lieb-Robinson bound (or even an algebraic spreading of correlation which occurs in systems with power-law interactions), and $2)$ algebraic scaling of infinite-time value of the OTOC with the system size.
We also demonstrated that any local time-independent many-body system which follows ETH and with a local conserved quantity satisfies these two conditions. For local time-independent systems, the conservation of energy is sufficient to guarantee the algebraic scaling of the infinite-time value of OTOC with the system size, and the Lieb-Robinson bound is inherited from the boundness and locality of the Hamiltonian. This means that slow relaxation of OTOCs is typical in local time-independent systems, and by this we mean that it would typically occur with a time-independent Hamiltonian with single-site and two-site random terms.
For number-conserving systems, we show that one can predict an algebraic decay of the infinite-time OTOC with the system size. This is achieved by approximating the evolution of a large system in a particular number sector, with a time-dependent weighted average of systems in different number sectors. A key step in this derivation is that the OTOC for each number sector relaxes to a different value which we are able to predict analytically.
For random unitary circuits with local conserved quantities, in \cite{RakovszkyKeyserlingk2018, KhemaniHuse2018} the authors derived an algebraic relaxation of the OTOC via a hydrodynamic approach. Here we have provided a conceptually different approach which relies on locality and ETH, and it applies to deterministic, non-random, systems.
We stress that the results here presented are not due to a prethermalization dynamics because we consider, at each moment in time, the infinite-time OTOC which already gives the long-time dynamics after any possible prethermalization occurs. The dynamics, in fact, can also be slower than the algebraic relaxation we describe. A clear example would be the case of many-body localized systems \cite{GiamarchiSchulz1987, GiamarchiSchulz1988, BaskoAltshuler2006, AbaninPapic2017, AbaninSerbyn2019, NandkishoreHuse2015, Lee2019}.
There remain still open questions as, for instance, how (non-)local the dynamics can be for the OTOC to decay algebraically (although we note
that for power-law interactions one can still expect an algebraic relaxation), and what would be the corresponding hydrodynamic equations of motion for such systems. \\

{\bf Acknowledgements}: We are grateful to S. Gopalakrishnan, V. Khemani, D. Luitz, A. Nahum, M. Rigol and L. F. Santos for fruitful discussions. D. P. acknowledges support from the Ministry of Education of Singapore AcRF MOE Tier-II (Project No. MOE2018-T2-2-142).

\bibliography{bibliography}
\newpage
\appendix

\section{Variance of the OTOC evolution} \label{app:numerical_details}

\begin{figure}[!b]
\includegraphics[width=\columnwidth]{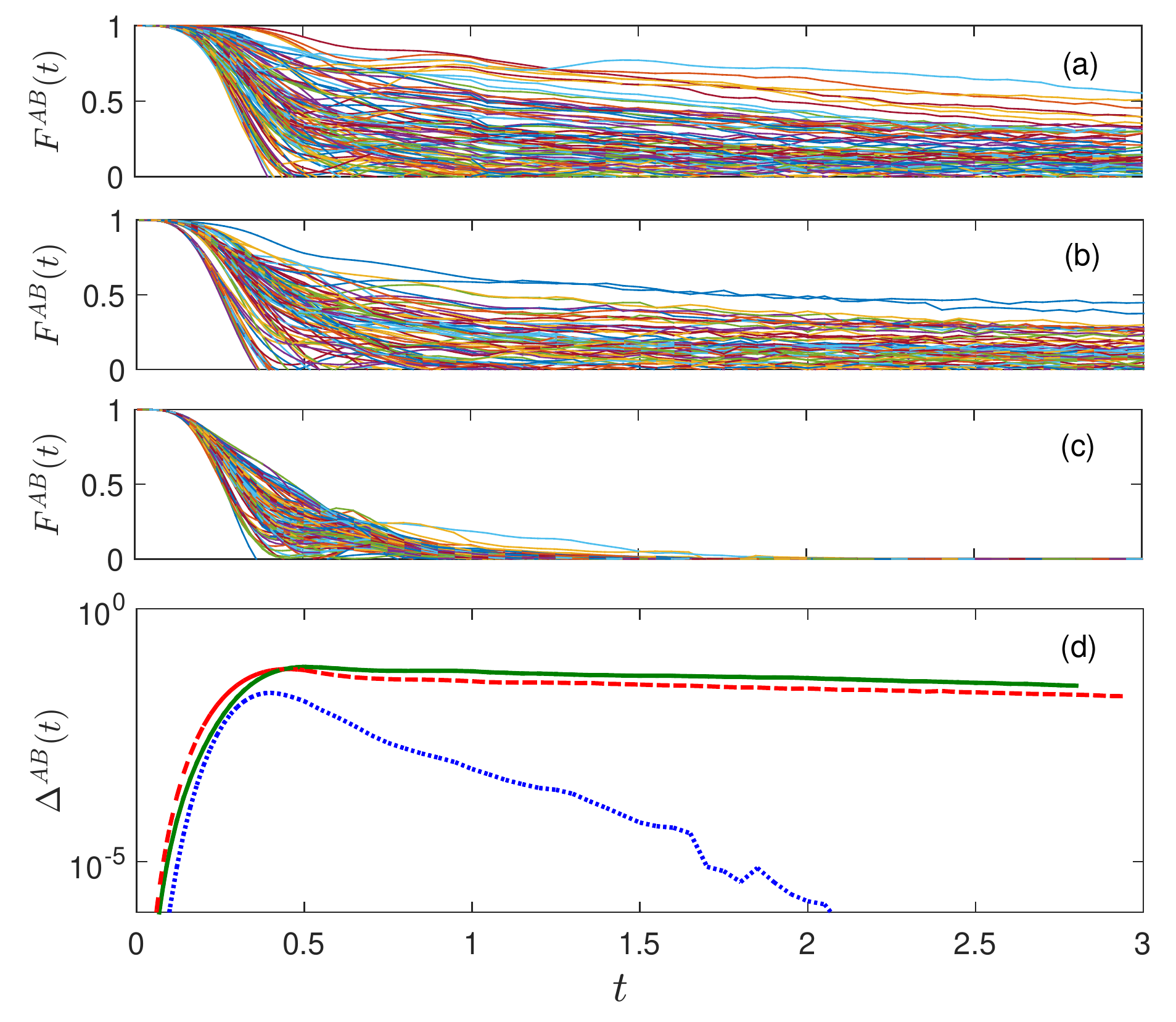}
\caption{(a)-(c) Time evolution of the OTOC $F^{AB}(t)$ of the non-integrable XXZ chain in Eq.(\ref{eq:XXZ_NNN}) with $J^z/J=2$, $\alpha=1$ and length $L=64$ in the zero magnetization sector for $100$ initial conditions.
In (a) $A=\sz_{L/2-1}$ and $B=\sz_{L/2+2}$, in (b) $A=\sz_{L/2-1}$ and $B=\sx_{L/2+2}$, and in (c) $A=\sx_{L/2-1}$ and $B=\sx_{L/2+2}$.
(d) Time evolution of variance of the OTOC $\Delta^{AB}(t)$ for the data in panel (a) green solid line, (b) red dashed line, and (c) blue dotted line. }
\label{fig:Fig2}
\end{figure}
We have shown that the finite time evolution of a number-conserving system is given by a weighted average of calculations done in different number
sectors, and also that the dynamics in each number sector may vary significantly. Hence, when computing the infinite-time OTOC from different initial conditions with different magnetization we can expect a significant deviation between each one of them.
In Fig.\ref{fig:Fig2}(a)-(c) we show $100$ different initial conditions chosen from different spin configurations of the chain. In panels (a) and (b) at least one operator is number-conserving and thus has nonzero diagonal elements in the energy basis. In these cases we observe a large spread of values between the OTOCs from different initial conditions. In panel
(c) instead, all trajectories tend to the same value, which is $0$, because the operators $A$ and $B$ do not have nonzero diagonal elements. In this case, after an initial time, the spread between the evolution of the different initial conditions is significantly suppressed.

To quantify this better, in Fig.\ref{fig:Fig2}(d) we study the time evolution of $\Delta^{AB}$, the variance of $F^{AB}$, given by
\begin{align}
\Delta^{AB}_{L,n}(t) = \langle \left(  A(t) B A(t) B - F^{AB}_{L,n}(t)\right)^2 \rangle_{L,n}.  \label{eq:variance}
\end{align}
What we observe is that the variance relaxes algebraically for the case in which at least one between $A$ or $B$ is number-conserving (green solid and red dashed lines), while it shows an exponential decay for the case in which neither $A$ nor $B$ is number-conserving and thus has zero diagonal elements in the energy eigenbasis (blue dotted line). Our numerical simulations confirms that the same applies to time-dependent number-conserving Hamiltonians.
\begin{figure}[!b]
\includegraphics[width=\columnwidth]{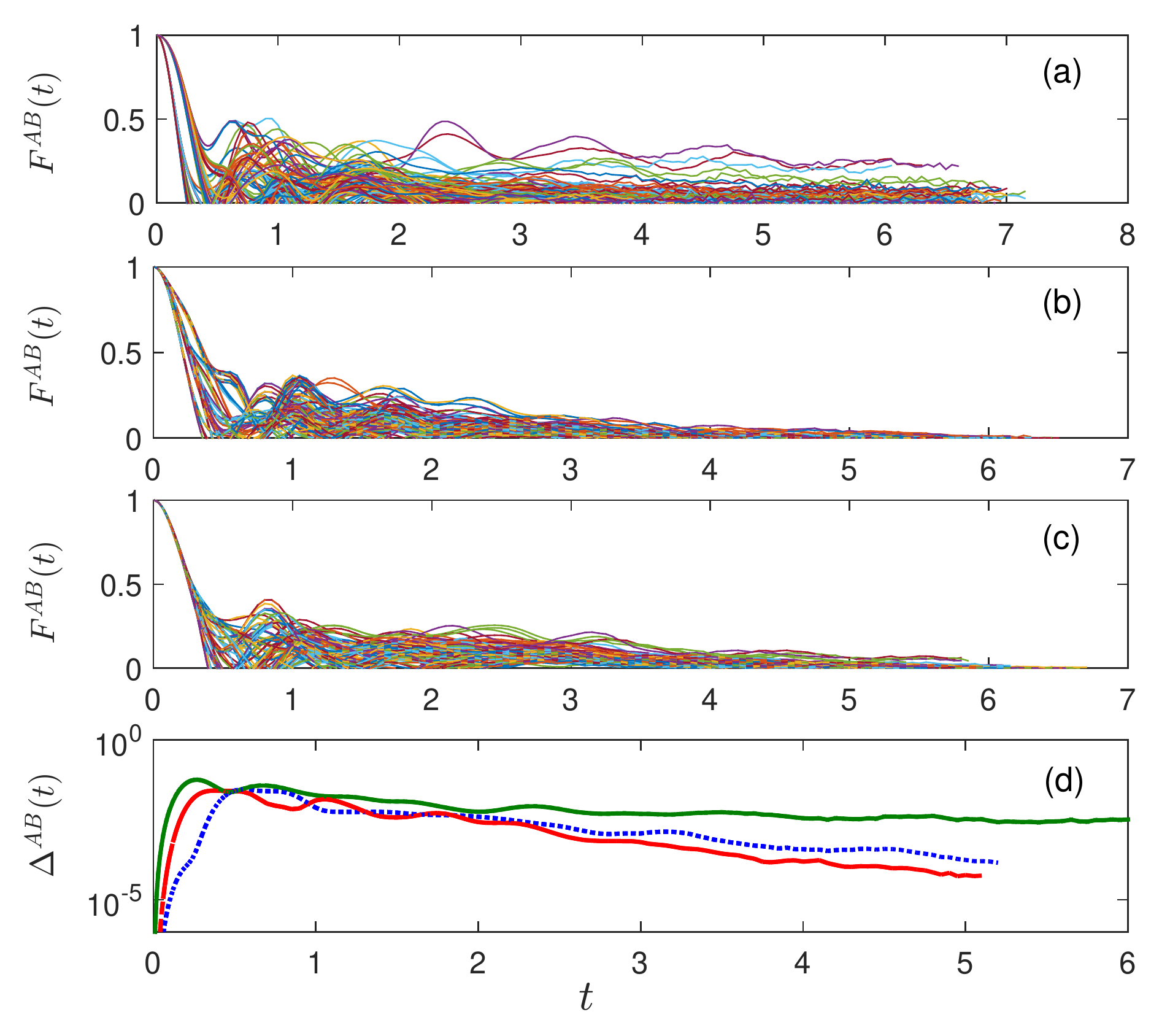}
\caption{(a)-(c) Time evolution of  OTOC $F^{AB}(t)$ of a non-integrable typical Hamiltonian in Eq. \ref{eq:typical} of length $L=30$ for $100$ initial conditions. Here we have used the randomly chosen parameters in Fig. \ref{fig:Fig1m}.
In (a) $A=\sz_{L/2}$ and $B=\sz_{L/2+1}$, in (b) $A=\sx_{L/2}$ and $B=\sx_{L/2+1}$, and in (c) $A=\sy_{L/2}$ and $B=\sy_{L/2+1}$.
(d) Time evolution of variance of the OTOC $\Delta^{AB}(t)$ for the data in panel (a) green solid line, (b) red solid line, and (c) blue dotted line. }
\label{fig:Fig7}
\end{figure}
The presence of a large variance of the OTOC in the case in which algebraic relaxation emerges is also observed when the only conserved quantity is energy, i.e. time-independent Hamiltonian. This is shown in Fig. \ref{fig:Fig7} for a random realization of the Hamiltonian in Eq. \ref{eq:typical} which is denoted by blue lines in  Fig. \ref{fig:Fig1m}. The time evolution of the OTOC $F^{AB}(t)$ for $100$ initial conditions is depicted in panels (a)-(c) with the corresponding variance $\Delta^{AB}(t)$ for these lines in bottom panels.  Top panel corresponds to time evolution of the OTOC $F^{AB}(t)$ with $A=\sz_{L/2}$, $B=\sz_{L/2+1}$ for which the overlap $\tr(HA)=tr(HB)\approx 0.8488$, whereas the panels (b) and (c) are  for $A=\sx_{L/2}$, $B=\sx_{L/2+1}$  and $A=\sy_{L/2}$, $B=\sy_{L/2+1}$ where the overlap is approximately  $0.1789$ and $0.0544$ respectively. Thus, we see a large variance when the overlap of the observable with Hamiltonian is large.

It is important to consider the larger variance in the OTOC for different
initial conditions in the case of algebraic relaxation, especially when studying large systems for which it is not practical to compute over all possible initial conditions. In fact in these cases one needs to sample over a large number of trajectories and consider the uncertainty of the values obtained.
In our computations we have considered a number of trajectories varying between $100$ to $500$, and for larger systems for which we used MPS we have considered bond dimensions $D=50,\; 100$, and we have checked that the results converged versus sampling size and that the standard error was small compared to the mean value.

\end{document}